\documentclass[journal]{IEEEtran}
\ifCLASSINFOpdf
\else
  \usepackage[dvips]{graphicx}
\fi
\begin{document}
%
\title{Modeling domain wall dynamics in thin magnetic strips with disorder}
%
%
%

\author{Lasse~Laurson$^{1}$,~
        Adil~Mughal$^{1}$,~
	Gianfranco~Durin$^{1,2}$,~
        and~Stefano~Zapperi$^{1,3}$\\
	\vspace{0.5cm}
	$^1$ISI Foundation, Viale S. Severo 65, 10133 Torino, Italy.\\
       	$^2$INRIM, Strada delle Cacce 91, 10135 Torino, Italy.\\
	$^3$CNR-INFM, S3, Universit\`a di Modena e Reggio Emilia, Via Campi 
	213/A, 41100 Modena, Italy.}

\maketitle

\begin{abstract}
We present a line-based model of transverse domain walls in 
thin magnetic strips, to study the effect of bulk disorder on the domain
wall dynamics within the thermally activated creep regime. The creep
velocity is found to exhibit a non-linear dependence on both applied 
magnetic fields and electric currents, characterized by similar creep
exponents for both forms of the external drive. We discuss briefly
the significance of the inherently stochastic thermally activated
domain wall motion from the point of view of spintronics applications,
where it generally is essential to be able to control the domain
wall displacement in a deterministic manner.
\end{abstract}


%
\IEEEpeerreviewmaketitle

\section{Introduction}
%
%
%
%
\IEEEPARstart{D}{omain} wall dynamics in thin magnetic strips or wires 
driven by either applied external magnetic fields or electric currents has
been an active field of research during recent years, due to both the
fundamental aspects of the underlying physics as well as due to the
potential spintronics applications \cite{BEA-08,PAR-08,BEA-05,ALL-05}. Most of the 
theoretical studies of such phenomena have so far focused on "perfect" systems 
free of any imperfections in the sample that could affect the dynamics of 
the domain wall. However, disorder is present in practically any 
realistic material, either in the form of edge roughness or various 
point-like defects in the bulk of the system. While the effect of edge
roughness \cite{MAR-072,YOS-03,BRY-05} or notches \cite{MAR-07} on domain 
wall motion has received much attention
in the literature, bulk disorder is usually assumed to be insignificant
in narrow nanostrips. However, this might not be true in general: for
instance, thickness fluctuations of the strip might give rise to
bulk disorder similarly to the disorder due to the rough edges of the 
system. Also other forms of impurities might be present in the bulk
of the system, affecting the domain wall dynamics.

In this paper we present a line-based model of a transverse domain
wall in a narrow and thin disordered magnetic (nano)strip. Such a coarse 
grained model is useful in understanding the effect of various kinds of 
disorder on the domain wall dynamics, because it does not suffer from the shortcomings
of some other approaches such as micromagnetic simulations where inclusion
of disorder is tricky due to mesh-related problems, or point-particle models
in which the internal degrees of freedom of the domain wall are not considered.
We focus on the effect of randomly distributed point-like pinning centers in the 
bulk of the system on the domain wall dynamics in the sub-threshold thermally
activated creep regime. In some recent studies, it has been demonstrated
that the experimentally relevant domain wall velocities often are
within this regime \cite{MAR-072}. The paper is organized as follows: in the 
next Section, the line based model of a transverse domain wall is presented, 
and some numerical results on the creep motion of such a domain wall are
presented in Section III. Section IV finishes the paper with discussion
and conclusions.


 

\section{Model}

\begin{figure}[!t]
\centering
\includegraphics[width=3.0in]{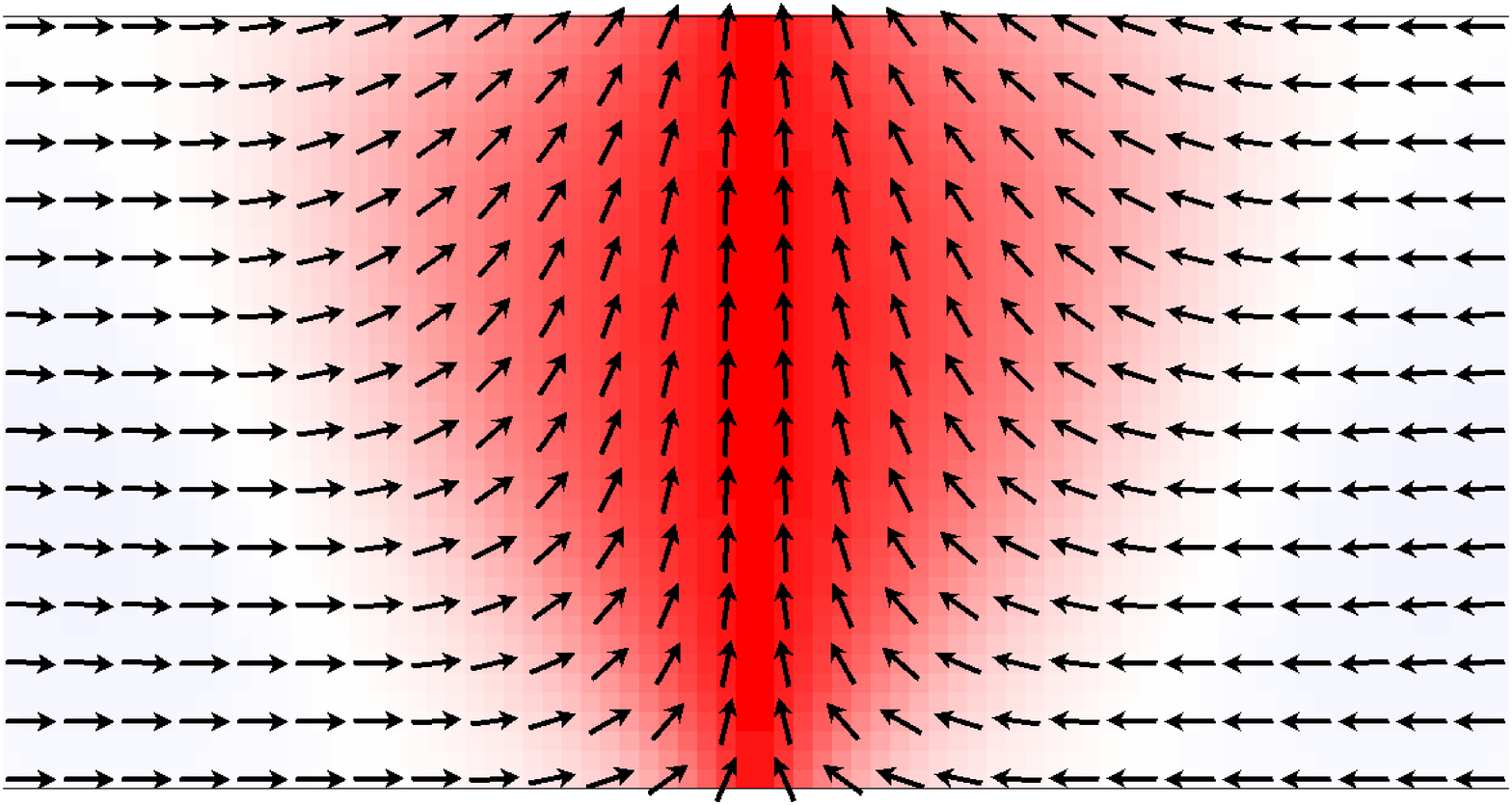}\\
\vspace{0.5cm}
\includegraphics[width=3.1in]{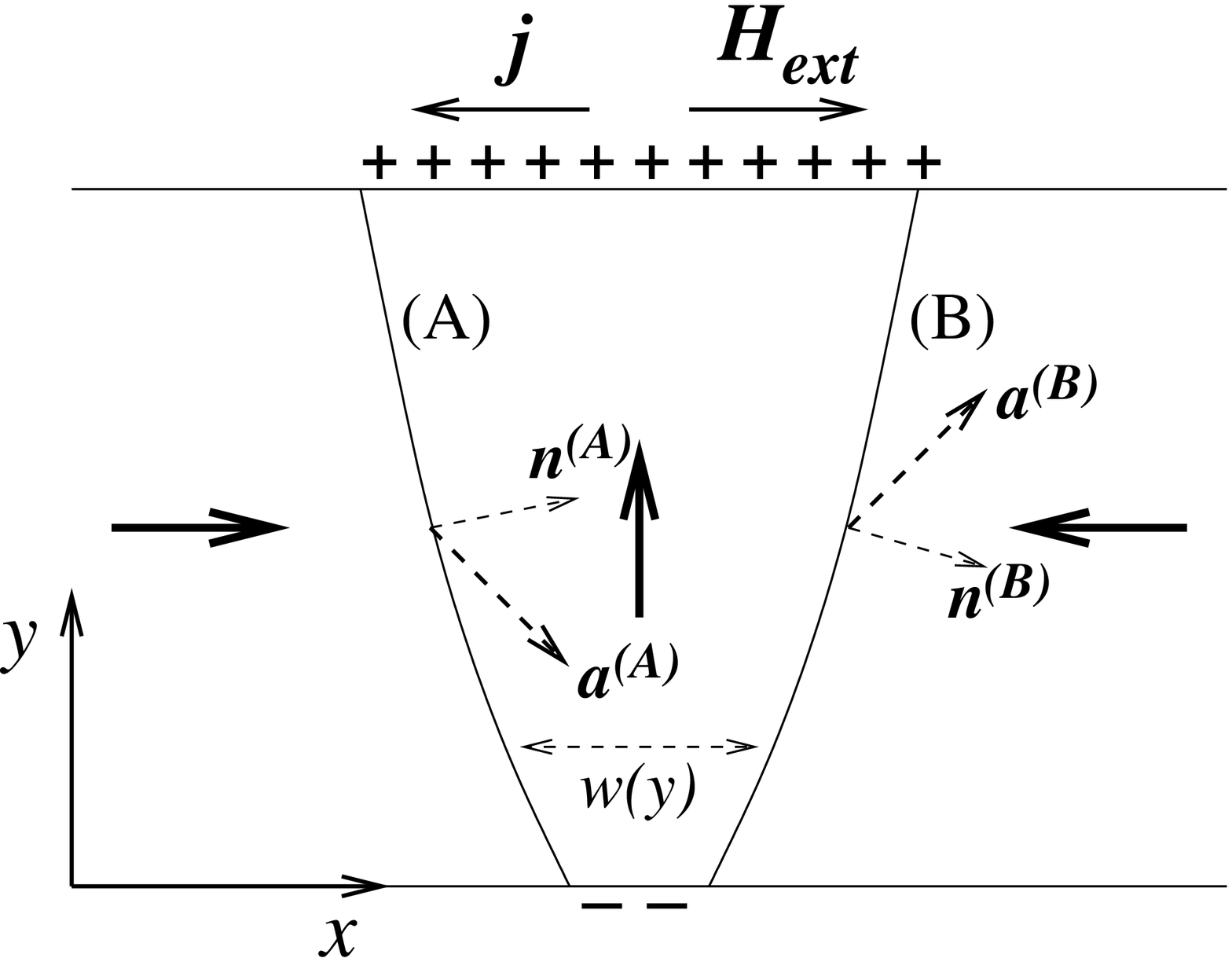}
\caption{{\bf Top:} An example of a transverse domain wall from micromagnetic 
simulations (using the OOMMF program \cite{OOMMF}), with the material 
parameters those of Permalloy, and strip dimensions $L=5 \cdot 10^{-6} m$,
$W=160 \cdot 10^{-9} m$ and $D=8\cdot 10^{-9} m$. 
{\bf Bottom:} A schematic of the line-based model,
where two lines (labeled (A) and (B), respectively) separating three different 
magnetic domains (with the magnetization direction indicated by the
thick solid arrows) form a model of the transverse wall. 
The dashed arrows indicate the directions $\vec{a}^{(i)}$ ($i=A,B$)  
of the components of the local field which drive the lines 
to the positive $x$ direction, as well as the line normals $\vec{n}^{(i)}$. 
}
\label{fig:tdw}
\end{figure}

We consider here strips of length $L$, width $W$ and thickness $D$, 
satisfying $L \gg W \gg D$, made of a soft magnetic material such as Permalloy.
The domain wall structure in such a strip depends on the balance between
exchange and anisotropy energies. The latter is here taken to be 
dominated by shape anisotropy, and thus the domains lie along the long axis
of the strip.  We focus on the case in which the width $W$ and thickness 
$D$ of the strip are sufficiently small, so that the stable domain wall 
structure is the so called transverse wall \cite{MCM-97,NAK-05,LAU-06}. 
For an example of its micromagnetic structure, see the top panel of Fig. 
\ref{fig:tdw}. For wider and/or thicker strips (not considered here), a domain 
wall with vortex topology would have a lower energy.

Due to the balance between exchange interactions and the effect of the 
demagnetizing fields $\vec{H}_{dm}$ arising from the 
"magnetic charges" within the transverse domain wall, the equilibrium 
shape of the wall resembles the letter V. These charges are associated 
to the discontinuities of the normal component of the magnetization. 
A simplified description of the wall structure
is obtained by considering the wall to be composed of {\em two} $90^{\circ}$ 
domain walls, with the first one from the left in Fig. 1 separating
domains with spins pointing to the right and up, respectively, while the
second $90^{\circ}$ wall separates the domain with spins up from from the
one with spins to the left. Clearly, such a domain description cannot be 
completely accurate as the magnetization within the wall rotates
smoothly from right to left as in Fig. \ref{fig:tdw}, so the 
magnetic charges associated to this rotation are actually distributed across the 
transition region. Nevertheless, by imposing the above approximation of 
two 90-degree walls with the magnetic charges concentrated along these walls yields a 
qualitatively correct equilibrium shape for the transverse wall, i.e.
the two 90-degree walls form a V-shaped structure.

The lower panel of Fig. \ref{fig:tdw} shows a schematic of the model. The two lines,
labeled (A) and (B) in the lower panel of Fig \ref{fig:tdw}, have line 
tensions $\gamma_w$, and interact repulsively due to the exchange interaction. 
With the local width of the wall $w(y)$, exchange interactions along the
strip axis lead to an energy $E_{e} \sim 1/w(y)$, resulting in a repulsive 
force $F_{e}(y) \sim 1/w(y)^2$ between any pair of line segments with the 
same $y$-coordinates. These forces are balanced by the forces due to the 
demagnetizing fields arising from the magnetic charges. The top edge 
(with $y=W$) has a positive charge density $\sigma=M_s$, where $M_s$ is the 
saturation magnetization. Similarly, the bottom edge (where $y=0$) has 
$\sigma=-M_s$. As described above, also the lines are charged: whenever the 
local angle between 
the line and the $x$-axis is different from 45 degrees, the lines have a 
non-zero charge density. Notice that the total 
charge of the domain wall is counterbalanced by the negative charges at the 
two ends of the strip (with $x=0$ and $x=L$). 

Furthermore, the domain wall can be driven towards the positive $x$ 
direction either by applying an external magnetic field 
$\vec{H}_{ext}=H_{ext}\hat{x}$ or an electric current $\vec{j}=-\jmath\hat{x}$
(with the direction chosen in such a way that the electrons
move along the $+\hat{x}$ direction). 
Here we restrict ourselves to the case of low fields and currents:
For stronger driving one should also include a "phase" angle of the domain
wall magnetization as a variable to the model. 
For the moment we assume that all magnetic moments remain within the plane 
of the strip. In other words, the fields and currents are restricted to be well below
the so called Walker threshold, where precession of the domain wall 
magnetization about the long axis of the strip would take place (something
that in the strip geometry considered here would in fact proceed via
nucleation and propagation of an {\em  antivortex}).
To see how the effect of a small current can be included in the present 
model, we point out that the Landau-Lifshitz-Gilbert (LLG) equation describing the
time evolution of the magnetization $\vec{M}(\vec{r},t)$, with the 
additional terms describing the effect of current can be written \cite{ZHA-04}
\begin{eqnarray}
\frac{\partial\vec{M}}{\partial t}& = & - \gamma \vec{M} \times \vec{H}_{eff}+
\alpha \vec{M}\times\frac{\partial \vec{M}}{\partial t}-
v_j\frac{\partial \vec{M}}{\partial x} +\\ \nonumber
& & + \beta v_j \vec{M}\times\frac{\partial \vec{M}}{\partial x},
\end{eqnarray}
where $\vec{H}_{eff}$ is the total effective magnetic field, $\gamma$ is the
gyromagnetic ratio and the second term is the Gilbert damping term. The effects 
due to an electric current are taken into account by the last two terms, 
usually referred to as the {\em adiabatic} and {\em non-adiabatic} terms, 
respectively. The essential point here is that the effect of current enters in both 
terms through the gradient of $\vec{M}$: The non-adiabatic term enters like a 
magnetic field that is proportional to $\partial \vec{M}/\partial x$, while
the adiabatic term exerts a torque on the spins of the domain wall, which 
is again proportional to $\partial \vec{M}/\partial x$. Thus, the
simplest way to include the effects of the current 
is to add a local magnetic field which
is proportional to the rate of change of the magnetization across the
domain wall, i.e. to $1/w(y)$. Thus, the difference between the two forms of 
driving is that whereas the effect of the field is independent of the local domain
wall width, the same is not true for the current drive.

The total magnetic field due to all these contributions is then 
evaluated at each segment of the discretized lines. While the domain wall
as a whole is driven towards the $+\hat{x}$ direction by a field along
the positive $x$ direction, the same is not true for the two ``sub-walls''
separately: Due to the different spin orientations on each side of the two
sub-walls, the relevant field components driving the sub-walls are also different
from each other. For the line 
(A), the relevant component of the local magnetic field driving the line 
towards the $+\hat{x}$ direction is the component along 
$\vec{a}^{(A)}=1/\sqrt{2}(\hat{x}-\hat{y})$, while the relevant component 
for line (B) is along $\vec{a}^{(B)}=1/\sqrt{2}(\hat{x}+\hat{y})$. 
Notice that the {\em total} field acting on the domain wall as a whole
is along the positive $x$ direction ($\vec{a}^{(A)}+\vec{a}^{(B)}=\sqrt{2} 
\hat{x}$). A force 
with a magnitude of these components of the local magnetic field is then 
taken to act on each segment of the lines along the local normal $\vec{n}_j$. 
This leads 
to a local normal force along each segment of the two lines, which is then used 
to calculate the dynamics of the system. To account for the effect of the 
finite damping coefficient in Eq. (3), we introduce a domain wall mass $m$, 
along with a friction force, so that the equation of motion
for the line segment $j$ reads
\begin{eqnarray}
\label{eq:eom}
m\frac{d\vec{v}_j}{dt} & = & -\chi \vec{v}_j + 
[-\gamma_w \kappa_j + F_{e,j}^{(i)} -(\nabla U_p)\cdot \vec{n}_j + \\ \nonumber
& & + (\vec{H}_{dm,j} + \vec{H}_{ext,j})\cdot \vec{a}^{(i)} + 
\eta_j]\vec{n}_j,
\end{eqnarray}
where $\chi$ is the effective friction coefficient, $\gamma_w$ is the line 
tension, $\kappa_j$ is the local curvature of the line and $\vec{n}_j$ 
is the local normal vector of the line. $F_{e,j}^{(A)}=-b/[w(y)]^2$ and 
$F_{e,j}^{(B)}=+b/[w(y)]^2$ are the repulsive forces between the two line 
segments of the different lines with the same $y$ coordinate due to exchange 
interactions. 
$U_p(\vec{r})=-C\sum_j e^{-\frac{1}{2}(\frac{\vec{r}-\vec{r}_j}{r_0})^2}$ is a 
quenched random potential describing the interaction of the line segments 
with various defects localized at random positions $\vec{r}_j$ within the 
strip (with $r_0$ the range of interaction). $\vec{H}_{dm,j}$ and $\vec{H}_{ext,j}$ are the demagnetizing
and external fields acting on the line segment $j$, respectively. As discussed
earlier, we include the effects due to an electric current as a local field
with a magnitude inversely proportional to $w(y)$, by setting
$\vec{H}_{ext,j}=[H_{ext}-\jmath/w(y)]\hat{x}$, where $H_{ext}$ is the magnitude
of the applied external field and $\jmath$ is the applied current density.
Finally thermal effects are included by a random force $\eta_j$ acting on each 
segment $j$ of the line, satifying $\langle \eta_j(t) \rangle=0$ and 
$\langle \eta_j(t)\eta_k(t')\rangle = k_b T \chi \delta_{j,k}\delta(t-t')$.

\section{Numerical results}

\begin{figure}[!t]
\centering
\includegraphics[width=0.75in]{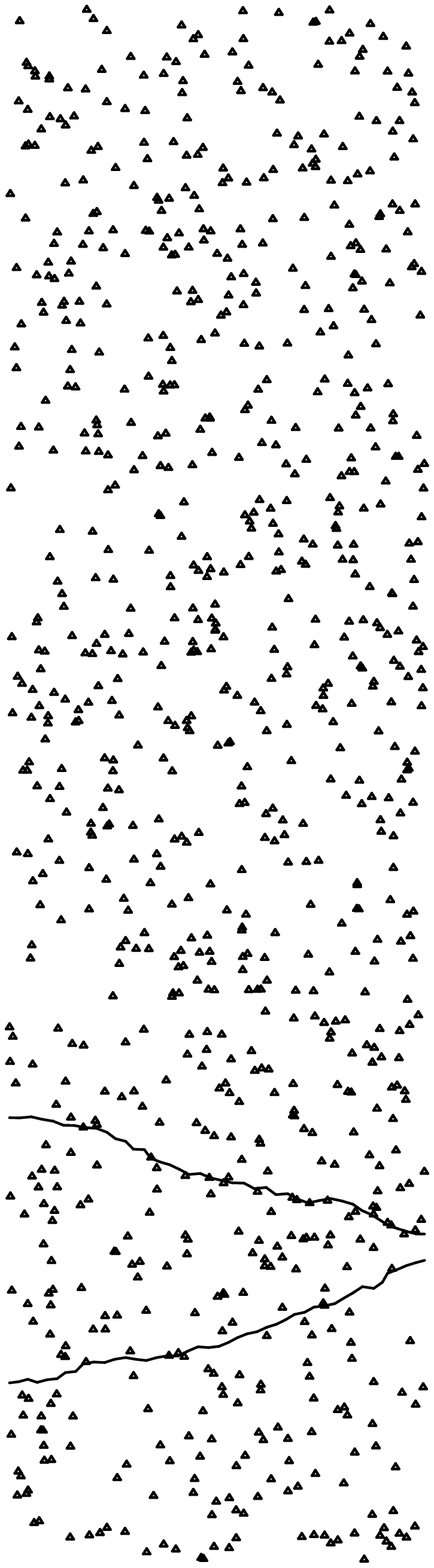}
\hspace{0.1in}
\includegraphics[width=0.75in]{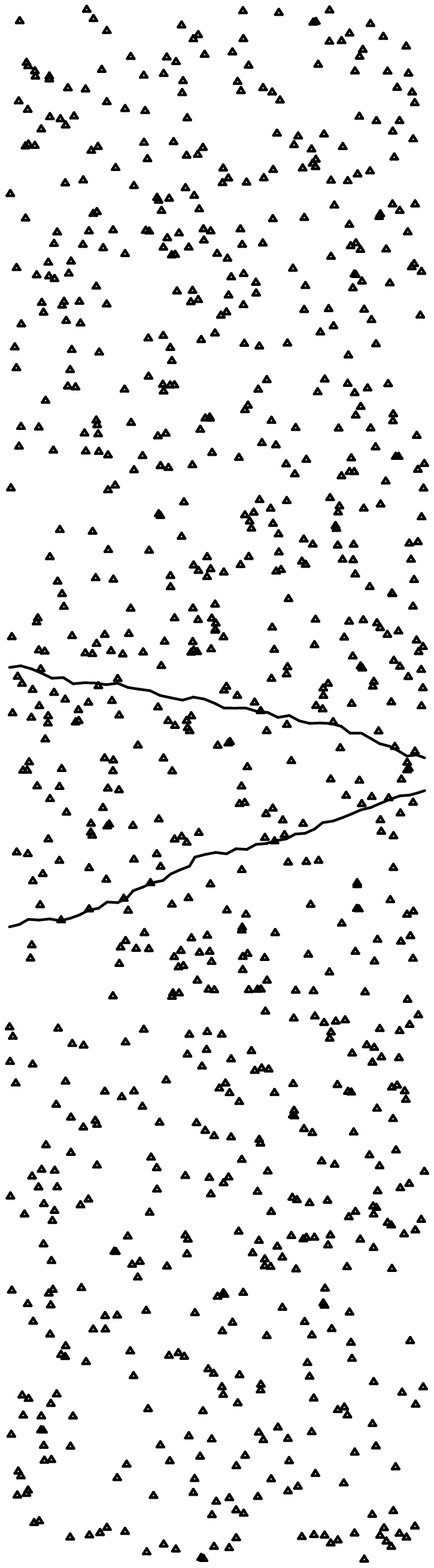}
\hspace{0.1in}
\includegraphics[width=0.75in]{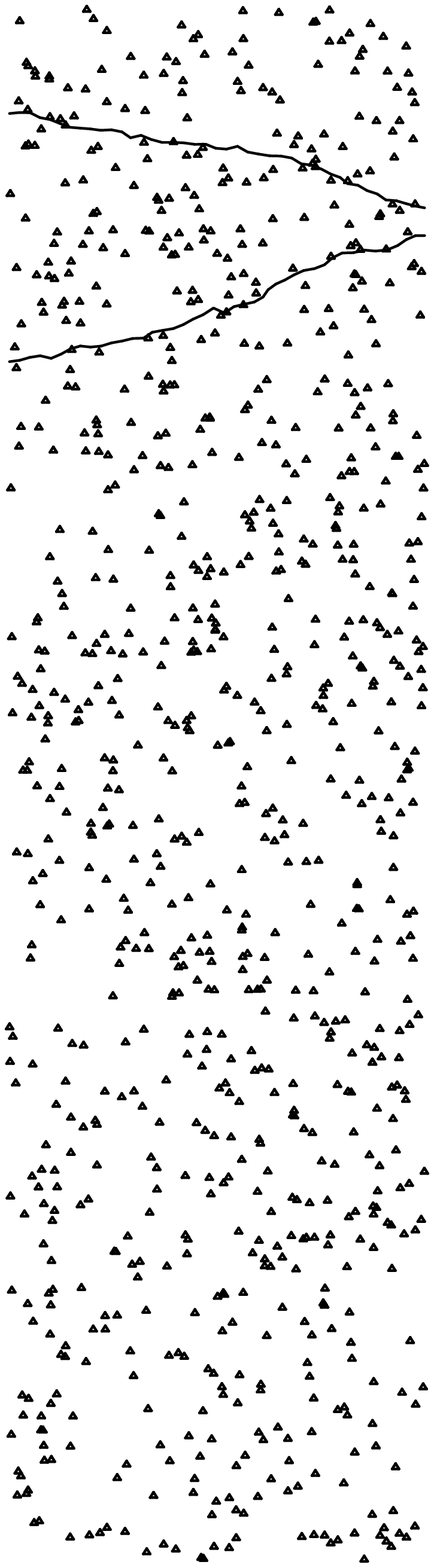}
\caption{Three snapshots for different times $t$ (from left to right, 
$t=5$, $t=17.5$ and $t=30$ in dimensionless units) of the domain wall moving 
in a disordered nanostrip (with the dots representing the pinning centers) under 
the influence of a subthreshold external field $H_{ext}$ and thermal fluctuations.}
\label{fig:snapshots}
\end{figure}

The model is implemented by using the software package Surface Evolver
\cite{se}, within which the line is taken to be formed by a set of vertices
connected to each other by edges. The program allows automatic remeshing of 
the discretized line, with the line moving in continuous space, thus avoiding 
spurious lattice effects. The equations of motion are integrated numerically 
with the Euler algorithm. In the absense of disorder, thermal effects and external
drive, the initial configuration of two parallel lines relax to reach 
a V-shaped stable structure, with the angle of the V-shape depending
on the relative magnitude of the demagnetizing fields and effects 
arising from the exchange interactions (the line tension of the lines,
the repulsive interaction between the two). After this initial
relaxation, the external drive, the interaction with disorder and thermal
effects are turned on and the evolution of the system is monitored. 
In the spirit of the model (where we imposed a restriction that the 
external driving is weak), we focus on the thermally activated
subthreshold creep regime. The parameters of the simulations in dimensionless
units of Eq. (\ref{eq:eom}) are as follows: 
$\chi=1$, $\gamma_w=1$, $b=0.025$, $m=0.001$, $C=0.2$ and $r_0=0.01$. 
The demagnetizing field due to a charge $M_s \sigma dl$ (with $\sigma$ measuring the
charge density in units of the saturation magnetization $M_s$) within a line segment
of length $dl$ positioned at the origin is taken have a magnitude $M_s \sigma dl/r^2$,
where the saturation magnetization is taken to have a value $M_s=0.15$, 
and $r$ is the distance from the origin. The unit of length is set by the 
sample dimensions, $L=4.5$ and $W=1.2$ (the very small thickness $D$ of the strip 
is neglected by treating the system as two-dimensional). Fig. \ref{fig:snapshots} 
shows an example of a sequence of configurations of the creeping domain wall 
in the field driven case.

\begin{figure}[!t]
\centering
\includegraphics[width=2.3in,angle=-90,clip]{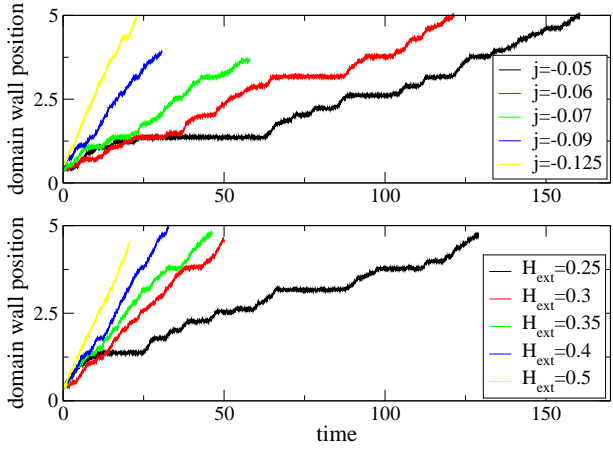}\\
\caption{Examples of the domain wall position along the long axis of the strip as a 
function of time for different magnitudes of the external drive within the creep 
regime. Top panel shows the current drive, while field drive is presented in the 
lower panel. Notice that the dynamics becomes increasingly erratic as the driving 
force is decreased.}
\label{fig:position}
\end{figure}

The dynamics of the domain wall in this regime displays the typical characteristics
of creep motion: The domain wall moves in discrete jumps, separated by periods
of pinning, see Fig. \ref{fig:position} for examples of the domain wall position 
as a function of time. Due to the thermally activated nature of the motion, the 
dynamics is inherently stochastic, and the relative fluctuations of the dynamics become
increasingly pronounced as the driving force is deacreased. 
 
\begin{figure}[!t]
\centering
\includegraphics[width=2.3in,angle=-90,clip]{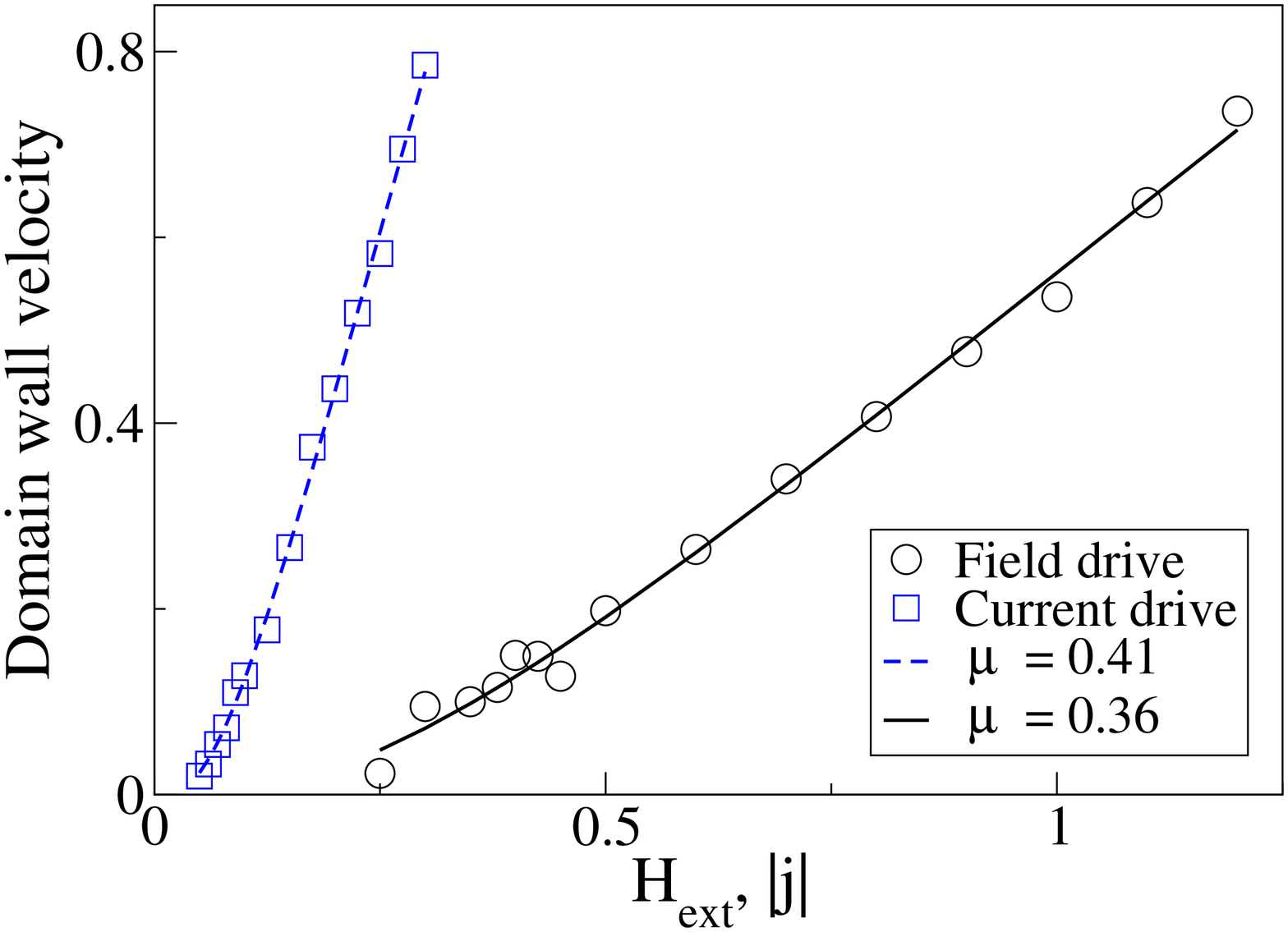}\\
\vspace{0.1in}
\includegraphics[width=2.37in,angle=-90,clip]{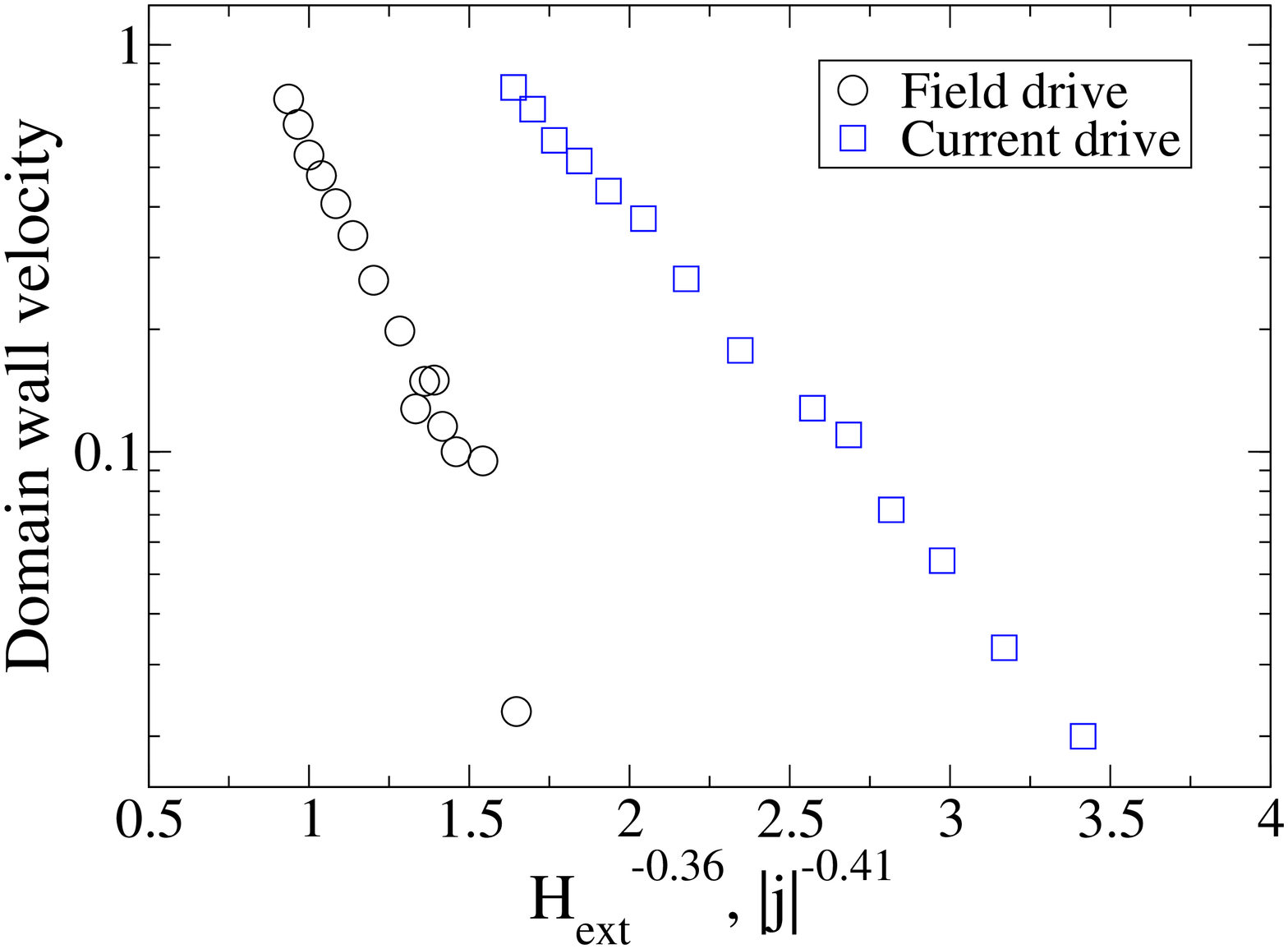}
\caption{The average creep velocity versus the external field $H_{ext}$ and 
current $\jmath$. The top panel shows the data (symbols) along with non-linear
fits of the form of Eq. (\ref{eq:creep}) (solid and dashed lines). The lower 
panel shows the same data scaled according to the fitted values of the
creep exponent $\mu$.}
\label{fig:velocity}
\end{figure}

The {\em average} creep velocity is found to exhibit non-linear dependence
on both external field and current. In general, for creep motion of
elastic manifolds subject to an applied force $f$ and temperature $T$, 
one expects the creep velocity to obey
\begin{equation}
\label{eq:creep}
v(f,T) \sim e^{-\frac{E_c}{k_b T}\left( \frac{f_c}{f}  \right)^{\mu} },
\end{equation}  
where $\mu$ is the creep exponent, characterizing the divergence of the height 
of the energy barriers as the driving force tends towards zero. Our  
results presented in Fig. \ref{fig:velocity} indicate that for both external 
field and current (i.e. $f=H_{ext}$ and $f=\jmath$), the data is consistent with 
the above form with $\mu=0.4 \pm 0.1$. A more detailed study of these issues will 
be published elsewhere. 

\section{Conclusions}

In this paper we have presented a line-based model of a transverse domain
wall in a disordered magnetic nanostrip, and studied its dynamics within
the thermally activated subthreshold creep regime. The creep motion of the
domain wall interacting with point-like pinning centers randomly positioned
within the bulk of the strip is found to exhibit typical features of creep
of elastic manifolds, i.e. stochastic velocity fluctuations and a non-linear
relation between the average creep velocity and the external drive. The
finding that the creep exponents for the field and current driven cases 
are similar indicates a universality of the two forms of driving in the
present case. This is in contrast with experiments on a different system 
with a different wall configuration
where different creep exponents were found in the two cases \cite{YAM-07}. 
Thus, a detailed experimental study of the creep dynamics of
transverse domain walls in nanostrips would be interesting, to see if the
simplified model presented here captures the statistical properties of 
the true dynamics. 
Notice also that the non-linear behaviour of the domain wall velocity versus 
the applied external drive is out of scope of any point particle based models, 
where one would expect $v(f) \sim f$.

Stochastic velocity fluctuations of domain wall motion in nanostrips have 
been observed also in experiments \cite{MEI-07,MOH-08,VAN-08}. This might suggest 
that these experiments are typically probing the creep regime, with the velocity
distribution arising from the random sequence of pinning and depinning events
governed by thermal fluctuations. From the point of view of practical
spintronics-based applications, this presents a fundamental problem: how to
displace a domain wall within a nanostrip in a controllable fashion? From the
point of view of low power consumptions of the devices, it might be desireable
to use low driving forces, but we have seen that within this regime the 
relative fluctuations in the domain wall dynamics are large. Using stronger
external drives leads to different problems, as the wall magnetization starts
to precess around the long axis of strip, something that in the presence of
disorder could also happen in a stochastic manner. As the translational and
phase degrees of freedom of the wall are coupled, this would presumably affect 
also the translational motion of the domain wall.

\section*{Acknowledgment}

LL wishes to thank Academy of Finland for financial support.

\ifCLASSOPTIONcaptionsoff
  \newpage
\fi


\begin{thebibliography}{1}
\bibitem{BEA-08}
G.~S.~D.~Beach, M.~Tsoi, and J.~L.~Erskine, J. Magn. Magn. Mater. {\bf 320} 1272
(2008).
\bibitem{PAR-08}
S.~S.~P.~Parkin, M.~Hayashi, and L.~Thomas, Science {\bf 320}, 190 (2008).
\bibitem{BEA-05}
G.~S.~.D.~Beach {\it et al.}, Nature Mater. {\bf 4}, 741 (2005); R.~Cowburn and
D. Petit, Nature Mat. {\bf 4}, 721 (2005).
\bibitem{ALL-05}
D.~A.~Allwood {\it et al.}, Science {\bf 309}, 1688 (2005).
\bibitem{MAR-072}
E.~Martinez {\it et al.}, Phys. Rev. B {\bf 75}, 174409 (2007).
\bibitem{YOS-03}
Y.~Nakatani, A.~Thiaville, and J.~Miltat, Nature Mat. {\bf 2} 521 (2003).
\bibitem{BRY-05}
M.~T.~Bryan, D.~Atkinson and R.~P.~Cowburn, J. Phys: Conf. Ser. {\bf 17}, 40 (2005).
\bibitem{MAR-07}
E.~Martinez {\it et al.}, Phys. Rev. Lett. {\bf 98}, 267202 (2007).
\bibitem{MCM-97}
R.~D.~McMichael {\it et al.}, IEEE Trans. Magn. {\bf 33} 4167 (1997).
\bibitem{NAK-05}
Y.~Nakatani, A.~Thiaville, J. Magn. Magn. Mater. {\bf 290-291}, 750 (2005).
\bibitem{LAU-06}
M.~Laufenberg {\it et al.}, Appl. Phys. Lett. {\bf 88}, 052507 (2006).
\bibitem{ZHA-04}
S.~Zhang and Z.~Li, Phys. Rev. Lett. {\bf 93}, 127204 (2004).
\bibitem{OOMMF}
M.~J.~Donahue and D.~G.~Porter, OOMMF User's Guide, Version 1.0, Interagency
Report NISTIR 6376, National Institute of Standards and Technology,
Gaithersburg, MD (Sept 1999).
\bibitem{se}
http://www.susqu.edu/brakke/evolver/evolver.html
\bibitem{YAM-07}
M.~Yamanouchi {\it et al.}, Science {\bf 317}, 1726 (2007).
\bibitem{MEI-07}
G.~Meier {\it et al.}, Phys. Rev. Lett. {\bf 98}, 187202 (2007).
\bibitem{MOH-08}
P.~M\"ohrke {\it et al.}, J. Phys. D {\bf 41}, 164009 (2008).
\bibitem{VAN-08}
A.~Vanhaverbeke, A.~Bischof, and R.~Allenspach, Phys. Rev. Lett. {\bf 101},
107202 (2008).

\end{thebibliography}
\end{document}